\documentstyle[psfig]{kapproc} 

\kluwerbib

\startauthorindex

\begin{document}

\articletitle{SACY - a Search for Associations Containing Young stars
\footnote{
Based on observations made under the
ON-ESO
agreement for the joint operation of
the 1.52\,m ESO telescope and at the  Observat\'{o}rio do Pico dos Dias
(LNA/MCT),  Brazil}
}
\author{ Carlos A. O. Torres, Germano R. Quast}
\affil{Laborat\'{o}rio Nacional de Astrof\'{\i}sica/MCT,
 37504-364 Itajub\'{a}, Brazil}
\email{beto@lna.br, germano@lna.br}
\author{Ramiro de la Reza,  Licio da Silva}
\affil{Observat\'{o}rio Nacional/MCT,
20921-030 Rio de Janeiro, Brazil}
\email{delareza@on.br, ldasilva@eso.org}
\author{Claudio H. F. Melo, Michael Sterzik}
\affil{ESO,  
Cassilla 19001 Santiago 19, Chile}
\email{cmelo@eso.org, msterzik@eso.org}

\chaptitlerunninghead{SACY}

\anxx{Torres et al.}

\begin{abstract}
The scientific goal of the SACY 
(Search for Associations Containing Young-stars) 
was to identify possible associations
of stars younger than the Pleiades Association
among optical counterparts of the
ROSAT X-ray bright sources.
High-resolution spectra for possible optical counterparts
later than G0 belonging to HIPPARCOS and/or TYCHO-2 catalogs
were obtained in order to assess both the
youth and the spatial motion of each target.
More than 1000 ROSAT sources were observed, covering a large
area in the Southern Hemisphere.
The newly identified young stars present a patchy distribution
in UVW and XYZ,
revealing the existence of huge nearby young associations.
Here we present the associations identified in this survey.

\end{abstract}


\section*{Introduction}
In 1989, de la Reza et al. searched for isolated T Tau stars (TTS)
and found a group of TTS around TW Hya.
This was the beginning of the Pico dos Dias survey (PDS). 
The PDS was a search for young stars using the IRAS Point
Source Catalog as the main selector (Gregorio-Hetem et al., 1992;
Torres et al., 1995; Torres, 1998).
X-ray sources  from the ROSAT All-Sky Survey (RASS)  gave a new
tool to find new young associations (Neuh\"{a}user 1997).
With some of these sources, Torres et al. (2000) found evidences for a
young nearby association: they  called it Horologium Association
(HorA). Almost simultaneously, Zuckerman \& Webb (2000) found
another one, very similar and adjacent in the sky, which they
called Tucana Association (TucA). In order to examine the physical
relation between both of them, and to search for other ones, we
started the SACY project (de la Reza et al. 2001; Torres et al.
2001; Quast et al. 2001).

\section{Observations}

For SACY we selected all bright RASS sources that
could be associated with
TYCHO-2 or HIPPARCOS stars with (B-V) $>$ 0.6, 
excluding well known
RS CVn, W UMa, giants, etc from  SIMBAD. 
We restricted our sample to stars later than G0
because we use the Li I 6707\AA\ equivalent width as
an age indicator.

We obtained high resolution spectra for the selected candidates with
the FEROS \'echelle spectrograph (Kaufer et al. 1999)
(resolution of 50000; spectral coverage of 5000\,\AA)
of the 1.52 m ESO telescope at La Silla
or  with the coud\'{e} spectrograph
(resolution of 9000; spectral coverage of 450\,\AA, centered at 6500\,\AA)
of  the 1.60\,m telescope of the Observat\'{o}rio do Pico dos Dias.
For some stars we obtained
radial velocities with  CORALIE at the Swiss Euler Telescope at ESO
(Queloz et al. 2000). We derived spectral classifications,
radial velocities and equivalent widths of {Li\,I 6707\,\AA} lines.
In particular, the Li\,I line is important since it can provide
a crude age estimate (Jeffries 1995) for late type stars.
If the Li\,I line equivalent width is larger than the
highest values for stars stars belonging to the Local
Association  (Neuh\"{a}user 1997), the star is  flagged as young.
This is shown in Figure 1. 
In Figure 2 we plot, in a polar projection, the complete sample observed.

\begin{figure}[ht]
\centerline{\psfig{file=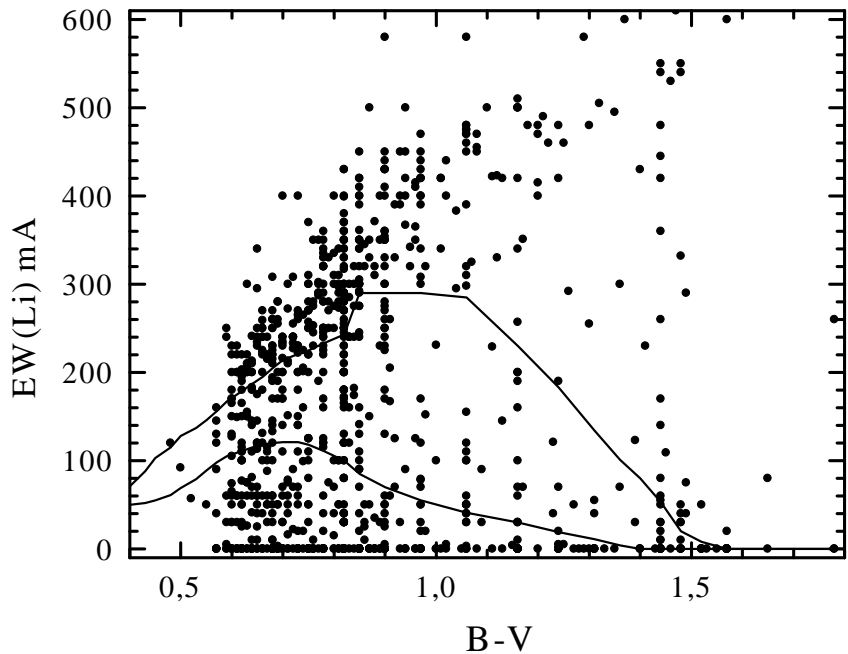,height=9cm}}
\caption{Distribution of Li line equivalent width for dwarf stars observed. 
The lines represent the upper and lower limits for Pleiades Association stars.}
\end{figure}

\begin{figure}[ht]
\centerline{\psfig{file=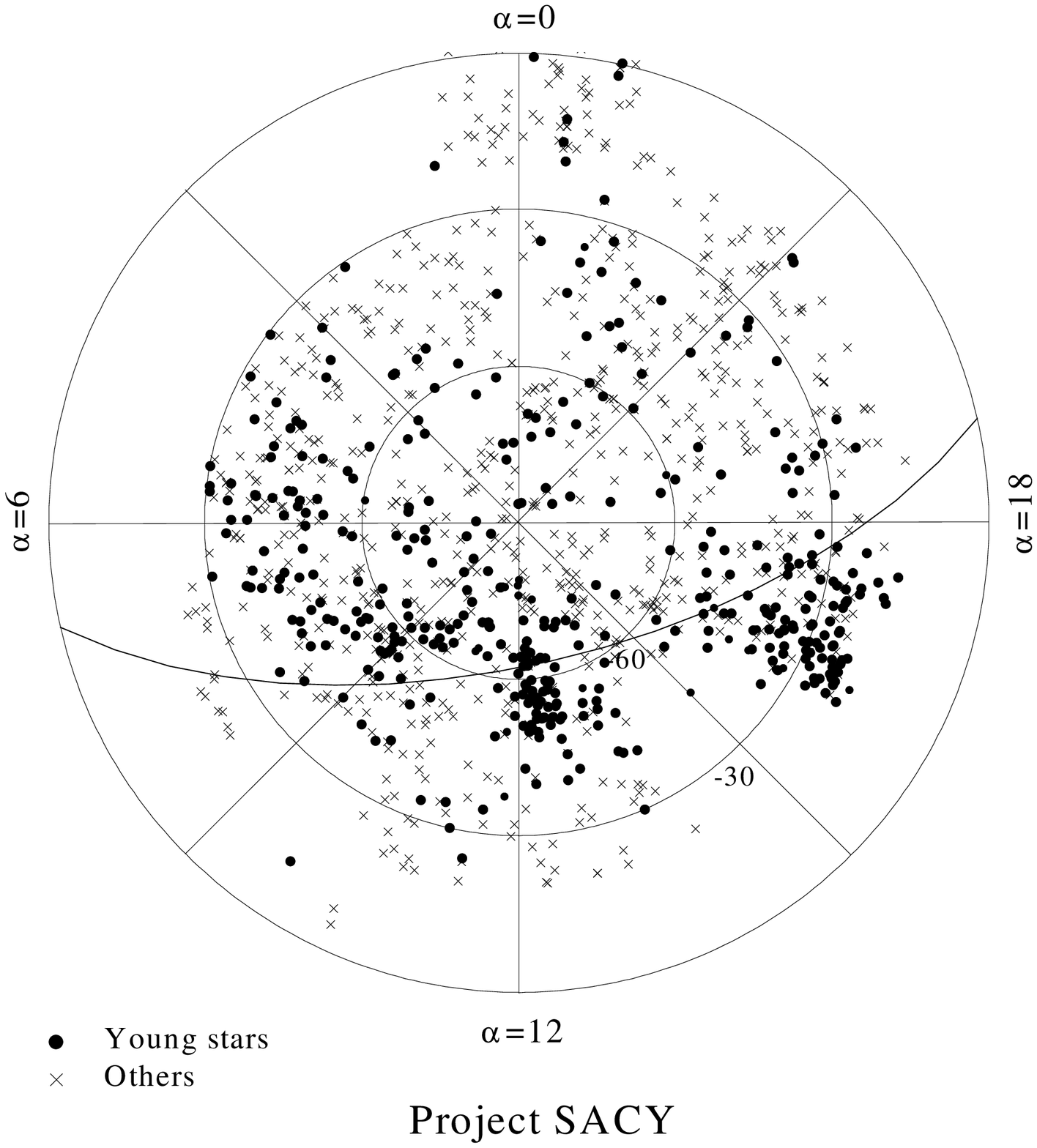,height=13cm}}
\caption{Polar representation of the observed stars where 
the transversal line is the galatic equator.}
\end{figure}

\subsection{Statistics}

There are 9574 RASS bright sources in the Southern Hemisphere,
2071 of them having counterparts with (B-V) $>$ 0.6 in TYCHO-2.
We observed 1096 sources. We also used published information for
99 others, most of them without interest for our search and
the young ones taken mainly from Covino et al. (1997).  
This defines the area in Figure 2, within which 
the SACY is complete to about 90\% .
Unfortunately  the southern area without observations is in the core
of the Sco-Cen Association.
We classified 201 stars as giants and 966 as dwarfs,
421 of them being younger than the Pleiades,
174 having the Pleiades age and 371 older than it.

\section{The Young Associations}

\begin{table}[ht]
\caption
{Space motions and parallaxes of the Young Associations}
\begin{tabular}{lrrrrrrrrr}
\sphline

Name&U&$\sigma$&V&$\sigma$&W&$\sigma$&$\pi$&$\sigma$&N$_*$\\
&km/s&km/s&km/s&km/s&km/s&km/s&mas&mas\\
\sphline
GAYA1&-9.1&1.1&-20.9&1.0&-1.2&0.9&22.0&2.2&16\\
GAYA2&-11.0&1.0&-22.5&1.0&-4.6&1.1&11.9&4.2&41\\
TWA&-12.1&0.7&-17.1&0.8&-5.5&0.9&22.4&5.6&5+3\\
$\epsilon$ ChaA&-7.7&0.6&-19.7&0.8&-8.5&0.9&10.5&1.3&15\\
LCC&-8.6&0.8&-21.4&1.1&-5.6&1.1&8.7&1.4&40\\
US&-4.7&1.2&-19.3&0.7&-5.2&1.3&7.5&1.4&43\\
YSSA&-4.0&1.3&-13.4&1.0&-8.0&1.2&8.6&1.9&21+5\\
$\beta$ PicA&-9.5&1.2&-16.4&1.1&-9.4&0.9&34.1&27.2&16+1\\
OctA&-10.4&0.6&-1.5&0.6&-8.0&0.8&8.9&1.5&6\\
ArgusA&-21.5&0.8&-13.1&1.1&-5.1&1.4&9.8&2.7&14\\
AnA&-7.1&0.8&-28.0&1.1&-12.4&1.2&19.0&10.5&11\\
\sphline

\end{tabular}

\end{table}
\begin{table} [ht]
\caption{Positions relative to the Sun and ages}
\begin{tabular}{lrrrrrrrrr}
\sphline

Name&X&$\sigma$&Y&$\sigma$&Z&$\sigma$&$\rho_{max}$&age&N$_*$\\
&pc&pc&pc&pc&pc&pc&pc&Myr&\\
\sphline

GAYA1&12&14&-25&8&-33&5&26&30&16\\
GAYA2&7&27&-78&33&-31&25&84&20&41\\
TWA&9&6&-41&10&20&5&18&8&8\\
$\epsilon$ ChaA&47&7&-82&12&-6&14&34&10&15\\
LCC&57&12&-100&20&16&17&75&15&40\\
US&132&27&-28&18&22&18&72&8&43\\
YSSA&117&22&-10&10&-2&30&50&8&26\\
$\beta$ PicA&33&28&-8&16&-16&7&50&15&17\\
OctA&65&30&-73&7&-55&6&46&30&6\\
ArgusA&30&30&-95&34&-10&25&83&30?&14\\
AnA&4&32&-35&28&-35&17&69&50&11\\
\sphline

\end{tabular}
\end{table}

Kinematical motions (UVW) have been found for the stars
measured by HIPPARCOS.
According to our age classification, we find some striking
concentrations in the young star UVW space that are not in the same
loci of the Pleiades age ones.
Older stars lack these concentrations.
This can be easily demonstrated with the mpeg-movies on the CD anex.
 
There are four main concentrations in UVW space for young stars 
that probably define distinct associations. 
In fact, two could be identified as previously
known associations: the Tuc-HorA -- that we call
GAYA (Great Austral Young Association) for its huge size --
and the $\beta$ Pic association. 
Kinematically and spatially near GAYA there is another
group of stars, but even more widespread and distant. 
The other association (AnA) is less characterized.    
 
Most of our young stars have no HIPPARCOS parallaxes, and we
applied the following kinematical analysis to find possible associations:
Each point in UVW  space  is taken as a convergence point and
we calculate for it the parallaxes of all stars such as to minimize the
moduli of the space velocity vectors relative to this point
(but, of course, preserving the parallaxes of HIPPARCOS stars).
Then we calculate the density of stars in the velocity space around
each point of a grid in UVW.
Around some points there are density concentrations much larger than
the background fluctuation, revealing possible associations.
All the main concentrations are also constricted in space, albeit
some of them cover large areas.

In Tables 1  \& 2 we present the properties of the  young
associations detected in this way. In Table 1 we give the mean
kinematical values and the mean parallax.
For some known associations we use
$bonafide$ members not observed in the SACY to help in their definition.
Their numbers are
indicated in  the last column in Table 1.
In Table 2 we give the mean XYZ, the age, 
and the distance ($\rho_{max}$)
of the most distant member with respect to the calculated center of
the association, giving an idea of its size.
The method gives no unique solution for
some stars, but in almost all cases we can infer a ``best 
membership".

\subsection{Comments about the  associations}

\subsubsection{GAYA1}  
We are calling GAYA (Torres et al. 2001)
two nearby concentrations on the UVW space,
separated mainly in W velocities. Both seem adjacent in the real
space. GAYA1 is somewhat older and is one of the more well 
defined of the associations
in SACY and the previous HorA and TucA are within it.  
Some of the proposed members  of TucA are outside of
the velocity definition (mainly the eastern ones).
Of their 16 proposed members 8 have parallaxes in HIPPARCOS.
The spread in distance is small and this does not seem an artifact either
of the SACY or our analysis as its derived center is at only 45\,pc. 
We tested 14 of its proposed members
for spectroscopic binarity without any positive detection.
GAYA1 seems deficient in binaries.
\subsubsection{GAYA2}
GAYA2 is much  less well defined, although it shows a clear concentration
using HIPPARCOS stars, but reinforced by members of Lower Cru-Cen (LCC). 
Actually, the UVW are very near the LCC ones 
and it is adjacent in space.
Nevertheless GAYA2 seems older than LCC and closer to us.

\subsubsection{TWA} The TW Hya association is not very well defined in SACY
since only two members have trigonometric parallaxes. 
Torres et al. (2003) present a list of proposed members, but many of them lack
information for a complete kinematical analysis. 
Anyway, we try to use all possible data. 
The convergence method has problems as TWA is near in velocity and
space of $\epsilon$ ChaA and LCC. 
We applied it limiting the possible spatial volume  but
including any star position in Torres et al. list. 
Nevertheless our solution excludes many stars in their list. 
We proposed as {\it bonafide} kinematical members: TWA 1, 2, 3, 4, 7, 8, 12
plus a new member, CD-39\,7538.
\subsubsection{$\epsilon$ ChaA}  
This association is defined by 
Mamajek, Lawson \& Feigelson (2000).
We propose new members, enlarging it.
There is ambiguity for about half of the proposed members
between $\epsilon$ ChaA and LCC, but the solutions show a
consistent separation in UVW space.
The distance found by us indicates it is in front of the Cha complex.
\subsubsection{LCC} This association has UVW near those of  $\epsilon$\,ChaA
and GAYA2 and the age seems between both. 
The LCC found in SACY is very similar to that found by Sartori et al. (2003)
for early-type stars.
\subsubsection{US} The Upper Sco (US) has UVW near those of LCC and YSSA. 
US can easily be separated from LCC in real space, but many stars 
may be assigned both to US and YSSA. 
Since we have almost no observations in Upper Cen-Lupus
we can not say if they would be separated in SACY.    
\subsubsection{YSSA}
This is a group of young stars, spread from $\rho$ Oph to R Cra,
with very similar properties, that we are now calling the
Young Sco-Sgr Association.
The western border engulfs the stars mentioned in  
Quast et al. (2001) and   Neuh\"{a}user  et al. (2000).
The split in space distribution can be explained by
the incompleteness in the RASS coverage.
Anyway, the convergence process gives some superposition with US association.
The distance is near the assumed one for the R CrA cloud.

\subsubsection{$\beta$ PicA}  
As described by Zuckerman et al.(2001) this association
is very close to the Sun.
We propose new members, some of then as far as 80\,pc, but the
distribution in space seems very consistent.
Among the new proposed members is V4046 Sgr, a notorious object,
classified before as an isolated SB classical TTS
(de la Reza et al., 1986; Quast, 1998; Quast et al. 2000).
WW PsA and TX PsA could be members (Song et al. 2002), 
but their parallaxes should be 49.5\,mas, closer than HIPPARCOS ones, 
about 2$\sigma$ of the HIPPARCOS errors.

\subsubsection{OctA}  
This is a very homogeneous small group of almost aligned stars
(all young G stars)  near the South Celestial Pole.
Since this region belongs to a completely surveyed area of the SACY,
new members have to be found by other means.
\subsubsection{ArgusA} Although not very well defined, it
has a special position in UVW. Since many stars are in Car, Vel and Pup
we tentatively propose to call it as Argus A.

\subsubsection{AnA} Like ArgusA, the main reason to claim
for this possible association are the very special UVW.
The majority of the proposed members 
have parallaxes and, therefore, this is one of the concentrations
in the HIPPARCOS sample.

GAYA1, GAYA2, LCC, US and YSSA form a decreasing sequence 
in age, going from west to east, 
and they seem to form a kind of continuum in UVW space.
All the associations but the last three in the 
tables can represent local aspects of a global star forming process.
More details of these associations can be seen in the poster
of Quast et al. in the enclosed CD. 
\begin{acknowledgments}
This work was partially supported by a CNPq - Brazil grant to C. A. O. Torres
(pr. 200356/02-0).

\end{acknowledgments}

\end{document}